\def\year{2019}\relax
\documentclass[letterpaper]{article} 
\usepackage{aaai19}  
\usepackage{times}  
\usepackage{helvet}  
\usepackage{courier}  
\usepackage{url}  
\usepackage{graphicx}  

\usepackage{subfigure}
\usepackage{amsmath}
\usepackage{amsfonts}
\usepackage{bbm}
\usepackage{multirow}
\usepackage{acronym}
\usepackage{enumitem}
\usepackage{booktabs}
\usepackage{xcolor}

\newcommand{\citet}[1]
{\citeauthor{#1}~\shortcite{#1}}
\newcommand{\citep}{\cite}

\acrodef{RNN}{recurrent neural network}
\acrodef{GRU}{gated recurrent unit}
\acrodef{MDP}{Markov decision process}

\begin{document}
%
\title{RepeatNet: A Repeat Aware Neural Recommendation Machine\\ for Session-based Recommendation}
\author{Pengjie Ren$^{1,2}$, Zhumin Chen$^1$, Jing Li$^1$, Zhaochun Ren$^3$, Jun Ma$^1$, Maarten de Rijke$^2$\\
$^1$ Shandong University, Jinan, China\\
$^2$ University of Amsterdam, Amsterdam, The Netherlands\\
$^3$ Data Science Lab, JD.com, Beijing, China\\
}
\maketitle
\begin{abstract}
Recurrent neural networks for session-based recommendation have attracted a lot of attention recently because of their promising performance.
\textit{repeat consumption} is a common phenomenon in many recommendation scenarios (e.g., e-commerce, music, and TV program recommendations), where the same item is re-consumed repeatedly over time.
However, no previous studies have emphasized \textit{repeat consumption} with neural networks.
An effective neural approach is needed to decide when to perform repeat recommendation.
In this paper, we incorporate a repeat-explore mechanism into neural networks and propose a new model, called RepeatNet, with an encoder-decoder structure.
RepeatNet integrates a regular neural recommendation approach in the decoder with a new repeat recommendation mechanism that can choose items from a user's history and recommends them at the right time.
We report on extensive experiments on three benchmark datasets. 
RepeatNet outperforms state-of-the-art baselines on all three datasets in terms of MRR and Recall. 
Furthermore, as the dataset size and the repeat ratio increase, the improvements of RepeatNet over the baselines also increase, which demonstrates its advantage in handling repeat recommendation scenarios.
\end{abstract}


\section{Introduction}
Session-based recommendations have received increasing interest recently, due to their broad applicability in many online services (e.g., e-commerce, video watching, music listening) \citep{ijcai2017-511}.
Here, a session is a group of interactions that take place within a given time frame. 
Sessions from a user can occur on the same day, or over several days, weeks, or even months \citep{Quadrana:2017:PSR:3109859.3109896}.

Conventional recommendation methods tackle session-based recommendations based on either the last interaction or the last session.
\citet{Zimdars:2001:UTD:647235.720264} and \citet{Shani:2005:MRS:1046920.1088715} investigate how to extract sequential patterns to predict the next item using Markov models.
Then, \citet{Chen:2012:PPV:2339530.2339643} propose logistic Markov embeddings to learn the representations of songs for playlist prediction.
A major issue for these models is that the state space quickly becomes unmanageable when trying to include all possible sequences of potential user selections over all items.
Recurrent neural networks (\acsp{RNN})\acused{RNN} have recently been used for the purpose of session-based recommendations and attracted significant attention.
\citet{Hidasi2016} introduce \acp{RNN} with \acp{GRU} for session-based recommendation.
They introduce a number of parallel \ac{RNN} (p-\ac{RNN}) architectures to model sessions based on both clicks and features (images and text) of clicked items \citep{Hidasi:2016:PRN:2959100.2959167}.
\citet{Quadrana:2017:PSR:3109859.3109896} personalize RNN models with cross-session information transfer and devise a Hierarchical RNN model that relays and evolves latent hidden states of the RNNs across user sessions.
\citet{Li:2017:NAS:3132847.3132926} introduce an attention mechanism into session-based recommendations and outperform \citep{Hidasi2016}.
Though the number of studies that apply deep learning to session-based recommendation is increasing, none has emphasized so-called \textit{repeat consumptions}, which are a common phenomenon in many recommendation scenarios (e.g., e-commerce, music, and TV program recommendations), where the same item is re-consumed repeatedly over time.

\textit{Repeat consumption} exists because people have regular habits.
For example, we all buy the same things repeatedly, we eat at the same restaurants regularly, we listen to the same songs and artists frequently \citep{Anderson:2014:DRC:2566486.2568018}.
Table \ref{intro-table-1} shows the repeat consumption ratio for three benchmark datasets that are commonly used in related studies \citep{Hidasi2016,Li:2017:NAS:3132847.3132926}.
\begin{table}[h]
\centering
\caption{Repeat ratio (\%) on three benchmark datasets.}
\label{intro-table-1}
\begin{tabular}{lccc}
\toprule
\textbf{Datasets} & \textbf{Train} & \textbf{Validation} & \textbf{Test} \\ 
\midrule
YOOCHOOSE 1/4     & 25.52          & 25.51               & 26.02         \\ 
DIGINETICA        & 19.94          & 20.06               & 20.49         \\ 
LASTFM            & 20.72          & 20.42               & 20.95         \\ 
\bottomrule
\end{tabular}
\end{table}
Repeat consumption not only exists but also accounts for a large proportion of the interactions in some applications.
In this paper, we investigate \textit{repeat consumption} by incorporating a repeat-explore mechanism into neural networks and propose a new model called RepeatNet with an encoder-decoder structure.
Unlike existing work that evaluates a score for each item using a single decoder, RepeatNet evaluates the recommendation probabilities of each item with two decoders in a \textit{repeat mode} and an \textit{explore mode}, respectively.
In the \textit{repeat mode} we recommend an old item from the user's history while in the \textit{explore mode} we recommend a new item.
Specifically, we first encode each session into a representation.
Then, we use a repeat-explore mechanism to learn the switch probabilities between repeat and explore modes.
After that, we propose a \textit{repeat recommendation decoder} to learn the probabilities of recommending old items in the \textit{repeat mode} and an \textit{explore recommendation decoder} to learn the probabilities of recommending new items under the \textit{explore mode}.
Finally, we determine the recommendation score for an item by combining the mode switch probabilities and the recommendation probabilities of each item under the two modes in a probabilistic way.
The mode prediction and item recommendation are jointly learned in an end-to-end back-propagation training paradigm within a unified framework.

We carry out extensive experiments on three benchmark datasets. 
The results show that RepeatNet outperforms state-of-the-art baselines on all three datasets in terms of MRR and Recall. 
Furthermore, we find that as the dataset size and the repeat ratio increase, the improvements of RepeatNet over the baselines also increase, which demonstrates its advantages in handling repeat recommendation scenarios.

To sum up, the main contributions in this paper are:
\begin{itemize}[nosep]
\item We propose a novel deep learning-based model named RepeatNet that takes into account the \textit{repeat consumption} phenomenon. To the best of our knowledge, we are the first to consider this in the context of session-based recommendation with a neural model.
\item We introduce a repeat-explore mechanism for session-based recommendation to automatically learn the switch probabilities between repeat and explore modes. Unlike existing works that use a single decoder, we propose two decoders to learn the recommendation probability for each item in the two modes.
\item We carry out extensive experiments and analyses on three benchmark datasets. The results show that RepeatNet can improve the performance of session-based recommendation over state-of-the-art methods by explicitly modeling \textit{repeat consumption}.
\end{itemize}


\section{Related Work}

We survey related work in two areas: session-based recommendations and repeat recommendations.

\subsection{Session-based recommendation}

Conventional methods for session-based recommendation are usually based on Markov chains that predict the next action given the last action.
\citet{Zimdars:2001:UTD:647235.720264} propose a sequential recommender based on Markov chains and investigate how to extract sequential patterns to learn the next state using probabilistic decision-tree models. 
\citet{Mobasher:2002:USN:844380.844798} study different sequential patterns for recommendation and find that contiguous sequential patterns are more suitable for sequential prediction task than general sequential patterns. 
\citet{Shani:2005:MRS:1046920.1088715} present a \ac{MDP} to provide recommendations in a session-based manner and the simplest \ac{MDP} boils down to first-order Markov chains where the next recommendation can simply be computed through the transition probabilities between items. 
\citet{Yap:2012:ENR:2260963.2260969} introduce a competence score measure in personalized sequential pattern mining for next-item recommendations. 
\citet{Chen:2012:PPV:2339530.2339643} model playlists as Markov chains, and propose logistic Markov embeddings to learn the representations of songs for playlists prediction. 
A major issue with applying Markov chains to the session-based recommendation task is that the state space quickly becomes unmanageable when trying to include all possible sequences of potential user selections over all items.

\acp{RNN} have proved useful for sequential click prediction \citep{Zhang:2014:SCP:2893873.2894086}.
\citet{Hidasi2016} apply \acp{RNN} to session-based recommendation and achieve significant improvements over conventional methods. 
They utilize session-parallel mini-batch training and employ ranking-based loss functions for learning the model. 
Later, they introduce a number of parallel \ac{RNN} (p-\ac{RNN}) architectures to model sessions based on clicks and features (images and text) of clicked items \citep{Hidasi:2016:PRN:2959100.2959167}; they propose alternative training strategies for p-RNNs that suit them better than standard training. 
\citet{Tan:2016:IRN:2988450.2988452} propose two techniques to improve the performance of their models, namely data augmentation and a method to account for shifts in the input data distribution. 
\citet{Jannach:2017:RNN:3109859.3109872} show that a heuristics-based nearest neighbor scheme for sessions outperforms the model proposed by \citet{Hidasi2016} in the large majority of the tested configurations and datasets. 
\citet{Quadrana:2017:PSR:3109859.3109896} propose a way to personalize \ac{RNN} models with cross-session information transfer and devise a Hierarchical \ac{RNN} model that relays end evolves latent hidden states of the \acp{RNN} across user sessions.
\citet{Li:2017:NAS:3132847.3132926} explore a hybrid encoder with an attention mechanism to model the user's sequential behavior and intent to capture the user's main purpose in the current session.

Unlike the studies listed above, we emphasize the \textit{repeat consumption} phenomenon in our models.

\begin{figure*}
 \centering
 \subfigure{
 \includegraphics[width=0.9\textwidth]{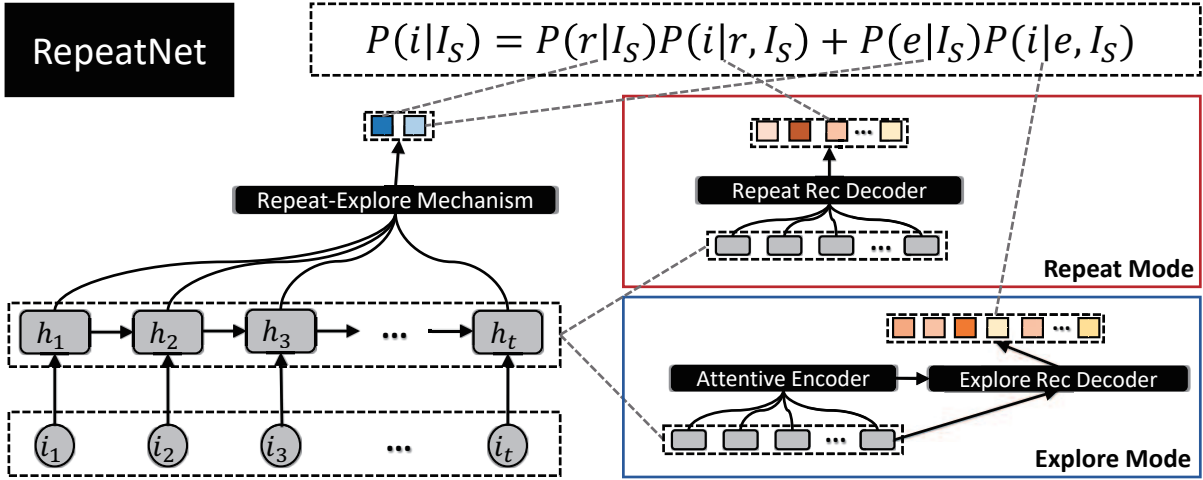}}
 \caption{Overview of RepeatNet.}
 \label{f_3_1}
\end{figure*}

\subsection{Repeat recommendation}

\citet{Anderson:2014:DRC:2566486.2568018} study the patterns by which a user consumes the same item repeatedly over time, in a wide variety of domains, ranging from check-ins at the same business location to re-watches of the same video.
They find that recency of consumption is the strongest predictor of repeat consumption.
\citet{Chen:2015:YRN:2887007.2887011} derive four generic features that influence people's short-term \textit{repeat consumption} behavior. 
Then, they present two fast algorithms with linear and quadratic kernels to predict whether a user will perform a short-term \textit{repeat consumption} at a specific time given the context.

An important goal of a recommender system is to help users discover new items.
Besides that, many real-world systems utilize lists of recommendation for a different goal, namely to remind users of items that they have viewed or consumed in the past.
\citet{Lerche:2016:VRW:2930238.2930244} investigate this through a live experiment, aiming to quantify the value of such reminders in recommendation lists.
\citet{Benson:2016:MUC:2872427.2883024} identify two macroscopic behavior patterns of repeated consumptions. First, in a given user's lifetime, very few items live for a long time. Second, the last consumptions of an item exhibit growing inter-arrival gaps consistent with the notion of increasing boredom leading up to eventual abandonment.
The main difference between our work and previous work on repeat recommendations is that we are the first to propose a neural recommendation model to explicitly emphasize \textit{repeat consumption} in both conventional and session-based recommendation tasks.


\section{RepeatNet}
Given an action (e.g., clicking, shopping) session $I_S=\{i_1,i_2,\ldots ,i_{\tau},\ldots ,i_t\}$, where $i_{\tau}$ refers to an item, session-based recommendation tries to predict what the next event would be, as shown in Eq.~\ref{conventional_methods}.
Without loss of generality, we take \emph{click actions} as our running example in the paper:
\begin{equation}
\label{conventional_methods}
P(i_{t+1}\mid I_S) \sim f(I_S),
\end{equation}
where $P(i_{t+1}\mid I_S)$ denotes the probability of recommending $i_{t+1}$ given $I_S$.
Conventional methods usually model $f(I_S)$ directly as a discriminant or probability function.


\subsection{Framework}

We propose RepeatNet to model $P(i_{t+1}\mid I_S)$ from a probabilistic perspective by explicitly taking \textit{repeat consumption} into consideration, as shown in Eq.~\ref{our_method}:
\begin{equation}
\begin{split}
P(i_{t+1}\mid I_S) = {}& P(r\mid I_S)P(i_{t+1}\mid r,I_S) + {}\\
&\quad P(e\mid I_S)P(i_{t+1}\mid e,I_S),
\end{split}
\label{our_method}
\end{equation}
where $r$ and $e$ denote \textit{repeat mode} and \textit{explore mode}, respectively.
Here, $P(r\mid I_S)$ and $P(e\mid I_S)$ represent the probabilities of executing in \textit{repeat mode} and \textit{explore mode}, respectively.
$P(i_{t+1}\mid r,I_S)$ and $P(i_{t+1}\mid e,I_S)$ refer to the probabilities of recommending $i_{t+1}$ in \textit{repeat mode} and in \textit{explore mode}, respectively, given $I_S$.

As illustrated in Fig.~\ref{f_3_1}, 
RepeatNet consists of four main components, a \textit{session encoder}, a \textit{repeat-explore mechanism}, a \textit{repeat recommendation decoder}, and an \textit{explore recommendation decoder}.
The \textit{session encoder} encodes the given session $I_S$ into latent representations $H=\{h_1,h_2$, $\ldots, h_{\tau},\ldots ,h_t\}$, where $h_t$ represents the session representation at timestamp $t$.
The \textit{repeat-explore mechanism} takes $H$ as input and predicts the probabilities of executing \textit{repeat mode} or \textit{explore mode}, corresponding to $P(r\mid I_S)$ and $P(e\mid I_S)$ in Eq.~\ref{our_method}.
The \textit{repeat recommendation decoder} takes $H$ as input and predicts the repeat recommendation probabilities over clicked items in $I_S$, corresponding to $P(i_{t+1}\mid r,I_S)$ in Eq.~\ref{our_method}.
The \textit{explore recommendation decoder} takes $H$ as input and predicts the explore recommendation probabilities over unclicked items in $I-I_S$, where $I$ refers to all items, corresponding to $P(i_{t+1}\mid e,I_S)$ in Eq.~\ref{our_method}.

\subsection{Session encoder}
Like previous studies \citep{Hidasi2016,Li:2017:NAS:3132847.3132926}, we use a GRU to encode $I_S$, where the GRU is defined as:
\begin{equation}
\begin{split}
&z_{\tau}=\sigma(W_z[emb(i_{\tau}),h_{{\tau}-1}]) \\
&r_{\tau}=\sigma(W_r[emb(i_{\tau}),h_{{\tau}-1}]) \\
&\widetilde{h_{\tau}}=\tanh(W_h[emb(i_{\tau}),r_{\tau}\odot h_{{\tau}-1}]) \\
&h_{\tau}=(1-z_{\tau})\odot h_{{\tau}-1} + z_{\tau}\odot \widetilde{h_{\tau}},
\end{split}
\end{equation}
where $W_z$, $W_r$, and $W_h$ are weight matrices;  $emb(i_{\tau})$ is the item embedding of $i_{\tau}$; $\sigma$ denotes the sigmoid function.
The initial state of the GRU is set to zero vectors, i.e., $h_0=0$.
After the \textit{session encoder}, each session $I_S$ is encoded into $H=\{h_1,h_2,\ldots ,h_{\tau},\ldots ,h_t\}$.

\subsection{Repeat-explore mechanism}
The \textit{repeat-explore mechanism} can be seen as a binary classifier that predicts the recommendation mode based on $H=\{h_1,h_2,\ldots ,h_{\tau},\ldots ,h_t\}$.
To this end, we first apply an attention mechanism \citep{Bahdanau2015} to $H$ to get a fixed-length vector representation of $I_S$.
Specifically, we first use the last hidden state $h_t$ to match with each encoder hidden state $h_{\tau} \in H$ to get an importance score:
\begin{equation}
e^{re}_{\tau}=v_{re}^{\top}\tanh(W_{re}h_t+U_{re}h_{\tau}),
\end{equation}
where $v_{re}$, $W_{re}$, and $U_{re}$ are parameters.
The importance scores are then normalized to get the context vector for $I_S$ as a weighted sum in Eq.~\ref{re-att}:
\begin{equation}
\label{re-att}
\begin{split}
\alpha^{re}_{\tau}={}&\frac{\exp(e^{re}_{\tau})}{\sum_{\tau=1}^{t}\exp(e^{re}_{\tau})}\\
c^{re}_{I_S}={}&\sum_{\tau =1}^{t}\alpha^{re}_{\tau} h_{\tau}.
\end{split}
\end{equation}

\noindent%
We then employ a softmax regression to transform $c^{re}_{I_S}$ into a mode probability distribution, corresponding to $P(r\mid I_S)$ and $P(e\mid I_S)$ respectively, as shown in Eq.~\ref{re}:
\begin{equation}
\label{re}
[P(r\mid I_S),P(e\mid I_S)]=\mathrm{softmax}(W^c_{re} c^{re}_{I_S}),
\end{equation}
where $W^c_{re}$ is the weight matrix.

\subsection{Repeat recommendation decoder}
The \textit{repeat recommendation decoder} evaluates the probability of re-clicking an item in $I_S$.
Inspired by CopyNet \citep{gu-EtAl:2016:P16-1}, we use a modification of the attention model to achieve this.
The probability of re-clicking item $i_{\tau} \in I_S$ is computed as follows:
\begin{align}
e^{r}_{\tau}={}&v_{r}^{\top}\tanh(W_{r}h_t+U_{r}h_{\tau}) \\
P(i\mid r, I_S)={}&
\begin{cases}
\frac{\sum_{i}{\exp(e^{r}_{\tau})}}{\sum_{\tau=1}^{t}\exp(e^{r}_{\tau})} & \text{if }i \in I_S\\
0 & \text{if }i \in I-I_S,
\end{cases}
\end{align}
where $v_{r}$, $W_{r}$, and $U_{r}$ are parameters;
$\sum_{i}{\exp(e^{r}_{\tau})}$ denotes the sum of all occurrences of item $i \in I_S$, because the same item might occur multiple times in different positions of $I_S$.

\subsection{Explore recommendation decoder}
The \textit{explore recommendation decoder} evaluates the probability of clicking a new item that does not exist in $I_S$.
To better capture the user's interest in session $I_S$, we employ an item-level attention mechanism that allows the decoder to dynamically select and linearly combine different parts of the input sequence \citep{Li:2017:NAS:3132847.3132926}:
\begin{equation}
\begin{split}
e^{e}_{\tau}={}&v_{e}^{\top}\tanh(W_{e}h_t+U_{e}h_{\tau}) \\
\alpha^{e}_{\tau}={}&\frac{\exp(e^{e}_{\tau})}{\sum_{\tau=1}^{t}\exp(e^{e}_{\tau})} \\
c^{e}_{I_S}={}&\sum_{\tau =1}^{t}\alpha^{e}_{\tau} h_{\tau},
\end{split}
\end{equation}
where $v_{e}$, $W_{e}$, and $U_{e}$ are parameters.
The factors $\alpha^{e}_h$ determine which part of the input sequence should be emphasized or ignored when making predictions.
We then combine the last hidden state and the attentive state into a hybrid representation $c_{I_S}$ for $I_S$, which is the concatenation of vectors $h_t$ and $c^{e}_{I_S}$:
$c_{I_S}=[h_t,c^{e}_{I_S}]$.

Finally, the probability of clicking item $i_{\tau} \in I-I_S$ is computed as follows:
\begin{align}
f_i={}&
\begin{cases}
-\infty & \text{if } i \in I_S \\
W^{c}_{e} c_{I_S} & \text{if } i \in I-I_S
\end{cases}
\\
P(i\mid e,I_S)={}&\frac{\exp(f_i)}{\sum_{\tau=1}^{t}\exp(f_\tau)},
\end{align}
where $W^{c}_{e}$ is the weight matrix and $-\infty$ means negative infinity.
Since $\exp(-\infty)=0$, we assume that if an item exists in $I_S$, then the probability of recommending it in the \textit{explore mode} is zero.

\subsection{Objective function}
Our goal is to maximize the output prediction probability given the input session. 
Therefore, we optimize the negative log-likelihood loss function as follows:
\begin{equation}
L_{rec}(\theta)=-\frac{1}{|\mathbb{I_S}|}\sum_{I_S \in \mathbb{I_S}}\sum_{\tau =1}^{|I_S|}\log P(i_{\tau}\mid I_S),
\end{equation}
where $\theta$ are all the parameters of RepeatNet,
$\mathbb{I_S}$ is the set of all sessions in the training set, and
$P(i_{\tau}\mid I_S)$ is the item prediction probability as defined in Eq.~\ref{our_method}.

RepeatNet incorporates an extra \textit{repeat-explore mechanism} to softly switch between \textit{repeat mode} and \textit{explore mode}.
We assume that if the next item exists in $I_S$, then it is generated under the \textit{repeat mode}, otherwise \textit{explore mode}.
Here, we can jointly train another mode prediction loss as follows, which is also the negative log-likelihood loss:
\begin{equation}
\begin{split}
&\mbox{}\hspace*{-7mm}L_{mode}(\theta) \\
={}&-\frac{1}{|\mathbb{I_S}|}\sum_{I_S \in \mathbb{I_S}}\sum_{\tau =1}^{|I_S|} \mathbbm{1}(i_{\tau} \in I_S)\log P(r\mid I_S) + {}\\
&\mbox{}\hspace*{25mm} (1-\mathbbm{1}(i_{\tau} \in I_S))\log P(e\mid I_S),
\end{split}
\end{equation}
where $\mathbbm{1}(i_{\tau} \in I_S)$ is an indicator function that equals 1 if $i_{\tau} \in I_S$ and 0 otherwise.

In the case of joint training, the final loss is a linear combination of both losses:
\begin{equation}
L(\theta)=L_{rec}(\theta)+L_{mode}(\theta).
\end{equation}

\noindent%
All parameters of RepeatNet as well as the item embeddings are learned in an end-to-end back-propagation training paradigm.
Due to the full probability term in Eq.~\ref{our_method}, the two modes probabilities $P(r\mid I_S)$, $P(e\mid I_S)$ and the item prediction probabilities $P(i\mid r,I_S)$, $P(i\mid e,I_S)$ are basically competing through a unified function.


\section{Experiments}

\subsection{Datasets and evaluation metrics}
We carry out experiments on three standard datasets, i.e., YOOCHOOSE, DIGINETICA, and LASTFM.
YOOCHOOSE and DIGINETICA are frequently used in session-based recommendation studies \citep{Hidasi2016,Tan:2016:IRN:2988450.2988452,Li:2017:NAS:3132847.3132926,Jannach:2017:RNN:3109859.3109872}.
Since they are both for e-commerce, we choose a third dataset in a different domain, music, Last.fm.\footnote{https://www.last.fm}
See Table~\ref{dataset}.
The splitting of the datasets are the same as \citep{Li:2017:NAS:3132847.3132926}.

\begin{itemize}[nosep]
\item YOOCHOOSE\footnote{http://2015.recsyschallenge.com/challenge.html} is a public dataset released by the RecSys Challenge 2015. 
We follow \citep{Hidasi2016,Li:2017:NAS:3132847.3132926} and filter out sessions of length 1 and items that appear less than 5 times. 
They note that the 1/4 version of the dataset is enough for the task and increasing the amount of data will not further improve the performance.
\item DIGINETICA\footnote{http://cikm2016.cs.iupui.edu/cikm-cup} is released by the CIKM Cup 2016. We again follow \citep{Li:2017:NAS:3132847.3132926} and filter out sessions of length 1 and items that appear less than 5 times.
\item LASTFM\footnote{http://www.dtic.upf.edu/~ocelma/MusicRecommendationDataset\\/lastfm-1K.html} is released by \cite{Celma:Springer2010} and widely used in recommendation tasks \citep{ijcai2017-511}. We use the dataset for music artist recommendation; we keep the top 40,000 most popular artists and filter out sessions that are longer than 50 or shorter than 2 items.
\end{itemize}

\noindent%
Recommender systems can only recommend a few items at a time, the actual item a user might pick should be amongst the first few items of the list \citep{8352808,Cheng:2018:ALF:3178876.3186145}. 
Therefore, commonly used metrics are MRR@20 and Recall@20 \citep{He:2018:APR:3209978.3209981,Mei:2018:AIN:3269206.3271813}.
In this paper, we also report MRR@10 and Recall@10.
\begin{itemize}[nosep]
\item Recall@k: The primary evaluation metric is Recall@k, which is the proportion of cases when the desired item is amongst the top-k items in all test cases. 
\item MRR@k: Another used metric is MRR@k (Mean Reciprocal Rank), which is the average of reciprocal ranks of the desire items. The reciprocal rank is set to zero if the
rank is larger than k. 
\end{itemize}

\begin{table}[t]
\centering
\caption{Statistics of three datasets (number of sessions and items).}
\label{dataset}
\begin{tabular}{l@{~~}r@{~~}r@{~~}r@{~~}r}
\toprule
\bf Dataset      & \bf Training & \bf Validation & \bf Test & \bf Items \\
\midrule
YOOCHOOSE & 5,325,971 & 591,775 & 55,898 & 30,470 \\
DIGINETICA    & 647,532 & 71,947 & 60,858 & 43,097 \\
LASTFM        & 2,690,424 & 333,537 & 338,115 & 40,000 \\
\bottomrule
\end{tabular}
\end{table}

\begin{table*}[h]
\centering
\caption{Experimental results (\%) on the three datasets.}
\label{results}
\begin{tabular}{@{~}l@{~~}cccccccccccc}
\toprule
\multirow{3}{*}{\bf Methods} & \multicolumn{4}{c}{\bf YOOCHOOSE}                        & \multicolumn{4}{c}{\bf DIGINETICA}                       & \multicolumn{4}{c}{\bf LASTFM}                           \\
\cmidrule(r){2-5}
\cmidrule(r){6-9}
\cmidrule(r){10-13}
                         & \multicolumn{2}{c}{MRR} & \multicolumn{2}{c}{Recall} & \multicolumn{2}{c}{MRR} & \multicolumn{2}{c}{Recall} & \multicolumn{2}{c}{MRR} & \multicolumn{2}{c}{Recall} \\
\cmidrule(r){2-3}
\cmidrule(r){4-5}
\cmidrule(r){6-7}
\cmidrule(r){8-9}
\cmidrule(r){10-11}
\cmidrule(r){12-13}
                         & @10        & @20        & @10          & @20         & @10         & @20       & @10          & @20         & @10         & @20       & @10          & @20         \\
\midrule
POP                      &      \phantom{0}0.26      &       \phantom{0}0.30     &      \phantom{0}0.81        &       \phantom{0}1.33      &     \phantom{0}0.18        &      \phantom{0}0.20     &       \phantom{0}0.53       &       \phantom{0}0.89      &        \phantom{0}1.09     &     \phantom{0}1.26      &     \phantom{0}2.90         &       \phantom{0}5.26      \\
S-POP                    &      17.70      &       17.79     &       25.96       &      27.11       &     13.64        &      13.68     &        20.56      &      21.06       &        \phantom{0}8.36     &     \phantom{0}8.71      &     18.08         &      22.59       \\
Item-KNN                 &      20.89      &      21.72      &        41.56      &      52.35       &        10.77     &     11.57      &       25.04       &      35.75       &      \phantom{0}4.48       &   \phantom{0}4.85        &     \phantom{0}9.77         &       14.84      \\
BPR-MF                   &      \phantom{0}1.90      &      \phantom{0}1.97      &       \phantom{0}3.07       &      \phantom{0}4.05       &       \phantom{0}1.86      &     \phantom{0}1.98      &       \phantom{0}3.60      &      \phantom{0}5.24       &      \phantom{0}4.88       &     \phantom{0}5.19      &       \phantom{0}9.87       &      14.05       \\
FPMC                     &      16.59      &       17.50     &        38.87      &      51.86       &      \phantom{0}6.30       &      \phantom{0}6.95     &       17.07       &       26.53      &       \phantom{0}4.58      &      \phantom{0}4.99     &        11.67      &     17.68     \\
PDP                     &      18.44      &       19.15     &        40.03      &      52.98       &      \phantom{0}6.75       &      \phantom{0}7.24     &       19.57       &       28.77      &       \phantom{0}4.86      &      \phantom{0}5.05     &        12.11      &     18.09     \\
\midrule
GRU4REC                  &      21.64      &      22.60      &       46.67       &     59.56        &      \phantom{0}7.59       &     \phantom{0}8.33      &       19.09       &      29.45       &      \phantom{0}4.92       &      \phantom{0}5.39     &       11.56       &      17.90       \\
Improved-GRU4REC         &     28.36       &      29.15      &       57.91       &      69.20       &       13.63      &     14.69      &       33.48       &      46.16
       &      \phantom{0}9.60       &      10.15     &       20.98       &        29.04     \\
GRU4REC-TOPK         &     29.76       &      30.69      &       58.15       &      70.30       &       13.14      &     14.16      &       31.54       &      45.23
       &      \phantom{0}7.44       &      \phantom{0}7.95     &       15.73       &        22.61     \\
NARM                     &      28.52      &      29.23      &       58.70       &      69.73       &       15.25      &     16.17      &       33.62       &       \textbf{49.70}      &       10.31      &    10.85       &      22.04        &       29.94      \\
\midrule
RepeatNet (no repeat)                     &      30.02      &      30.76      &       59.62       &      70.21       &       12.71      &     13.52      &       30.96       &       42.56      &       \phantom{0}9.92      &    10.47       &      21.81        &       29.96      \\
RepeatNet &      \textbf{30.50}\rlap{$^\dagger$}      &     \textbf{31.03}\rlap{$^\dagger$}     &      \textbf{59.76}\rlap{$^\dagger$}     &      \textbf{70.71}      &      \textbf{16.90}\rlap{$^\dagger$}       &      \textbf{17.66}\rlap{$^\dagger$}     &       \textbf{36.86}\rlap{$^\dagger$}       &       47.79      &      \textbf{11.46}\rlap{$^\dagger$}       &      \textbf{12.03}\rlap{$^\dagger$}     &      \textbf{24.18}\rlap{$^\dagger$}        &      \textbf{32.38}\rlap{$^\dagger$}       \\
\bottomrule
\end{tabular}
\begin{minipage}{\textwidth}
\small{\textbf{Bold face} indicates the best result in terms of the corresponding metric.
Significant improvements over the best baseline results are marked with $^\dagger$ (t-test, p $<$ .05). 
The scores reported in \citep{Li:2017:NAS:3132847.3132926} on the DIGINETICA dataset differ because they did not sort the session items according to the ``timeframe'' field, which ignores the sequential information.
We run the code released by \citep{Hidasi2016,Tan:2016:IRN:2988450.2988452,DBLP:journals/corr/HidasiK17,Li:2017:NAS:3132847.3132926} to obtain the results of GRU4REC, Improved-GRU4REC, GRU4REC-TOPK, and NARM. 
}
\end{minipage}
\end{table*}

\subsection{Implementation details}
We set the item embedding size and GRU hidden state sizes to 100. 
We use dropout \citep{Srivastava2014} with drop ratio $p$ = 0.5.
We initialize model parameters randomly using the Xavier method \citep{Glorot2010Understanding}. 
We use Adam as our optimizing algorithm. 
For the hyper-parameters of the Adam optimizer, we set the learning rate $\alpha = 0.001$, two momentum parameters $\beta1 = 0.9$ and $\beta2 = 0.999$, respectively, and $\epsilon$ = $10^{-8}$. 
We halve the learning rate $\alpha$ every 3 rounds.
We also apply gradient clipping \citep{Pascanu2013} with range $[-5, 5]$ during training. 
To speed up the training and converge quickly, we use mini-batch size 1024 by grid search.
We test the model performance on the validation set for every epoch.
The model is written in Chainer \citep{chainer_learningsys2015} and trained on a GeForce GTX TitanX GPU. 

\subsection{Methods used for comparison}

\subsubsection{Conventional methods}

We select the following conventional methods which are commonly used as baselines in session based recommendations \citep{Hidasi2016,Tan:2016:IRN:2988450.2988452,Li:2017:NAS:3132847.3132926}.

\begin{itemize}[nosep]
\item POP: POP always recommends the most popular items in the training set. It is frequently used as baselines in recommender system domains \citep{He:2017:NCF:3038912.3052569}.
\item S-POP: S-POP recommends the most popular items of the current session.  Ties are broken using global popularity values \citep{Hidasi2016}.
\item Item-KNN: Items similar to the actual item are recommended by this baseline. Similarity is defined as the co-occurrence number of two items in sessions divided by the square root of the product of the number of sessions in which either item occurs. Regularization is also included to avoid coincidental high similarities between rarely visited items \citep{Davidson:2010:YVR:1864708.1864770}.
\item BPR-MF: BPR-MF \citep{Rendle:2009:BBP:1795114.1795167} is a commonly used matrix factorization method. 
We apply it to session-based recommendation by representing a new session with the average latent factors of items that occurred in the session so far.
\item FPMC: FPMC \citep{Rendle:2010:FPM:1772690.1772773} is a state-of-the-art hybrid model for next-basket recommendation. To adapt it to session-based recommendation, we ignore the user latent representations when computing recommendation scores.
\item PDP: \citet{Benson:2016:MUC:2872427.2883024} propose PDP and claim that they are the first to model sequential repeat consumption. This is the only recommendation model that considers sequential repeat consumption, to the best of our knowledge.
\end{itemize}

\subsubsection{Deep learning methods}
No previous studies propose neural models that consider sequential repeat consumption.
We select recent state-of-the-art neural session based recommendation models as baselines.

\begin{itemize}[nosep]
\item GRU4REC: GRU4REC~\citep{Hidasi2016} uses session-parallel mini-batch training process and also employs ranking-based loss functions for learning the model. 
\item Improved-GRU4REC: Improved GRU4REC~\citep{Tan:2016:IRN:2988450.2988452} improves GRU4REC with two techniques, data augmentation and a method to account for shifts in the input data distribution.
\item GRU4REC-TOPK: \citet{DBLP:journals/corr/HidasiK17} further improve GRU4REC with a top-k based ranking loss.
\item NARM: NARM~\citep{Li:2017:NAS:3132847.3132926} further improves Improved-GRU4REC with a neural attention mechanism.
\end{itemize}


\section{Results and Analysis}

\subsection{Results}
The results of all methods are shown in Table~\ref{results}.
We run the code released by \citep{Hidasi2016,Li:2017:NAS:3132847.3132926} to report the results of GRU4REC and NARM.
We can get several insights from Table~\ref{results}.
First, RepeatNet outperforms both conventional methods and recent neural methods, including the strong baselines, GRU4REC-TOPK and NARM.
The improvement of RepeatNet over NARM is even larger than the improvement of NARM over Improved-GRU4REC.
The improvements mean that explicitly modeling \textit{repeat consumption} is helpful, which gives RepeatNet more capabilities to model complex situations in session-based recommendations.

Second, as the repeat ratio increases, the performance of RepeatNet increases generally.
We reach this conclusion based on the different improvements on YOOCHOOSE and DIGINETICA.
Both datasets are from the e-commerce domain but YOOCHOOSE has a higher repeat ratio.

Third, the performance of RepeatNet varies with different domains.
Table~\ref{results} shows that RepeatNet has a bigger advantage in the music domain than in the e-commerce domain;
we believe this is due to different characteristics of the different domains.
S-POP performs much better than Item-KNN on LASTFM, which means that popularity is very important on LASTFM.
However, Item-KNN performs much better than S-POP on YOOCHOOSE, which means that collaborative filtering is more important on YOOCHOOSE.
Besides, the neural models have substantial gains over the conventional methods in all evaluation metrics on all datasets generally.
Similar conclusions have been formulated in other recent studies \citep{Hidasi2016,Tan:2016:IRN:2988450.2988452,Li:2017:NAS:3132847.3132926}.



\subsection{Analysis of the repeat mechanism}

\begin{table}[h]
\centering
\caption{MRR@20 (\%) of RepeatNet (with and without repeat mechanism) on repeat and non-repeat sessions.}
\label{repeat_analysis1}
\begin{tabular}{llcc}
\toprule
\multicolumn{2}{c}{\bf RepeatNet}     & \bf With repeat & \bf No repeat \\ \midrule
\multirow{2}{*}{YOOCHOOSE} & Rep     &      58.78   &      \textbf{60.18}        \\  
                           & Non-Rep &     \textbf{21.60}     &      20.42        \\ \midrule
\multirow{2}{*}{DIGINTICA} & Rep     &      \textbf{56.27}    &      29.20        \\  
                           & Non-Rep &      \phantom{0}7.71    &    \textbf{\phantom{0}9.48}            \\ \midrule
\multirow{2}{*}{LASTFM}    & Rep     &      \textbf{41.63}       &       32.68      \\  
                           & Non-Rep &       \phantom{0}4.18     &   \textbf{\phantom{0}5.06}             \\ \bottomrule
\end{tabular}
\begin{minipage}{\columnwidth}
\small{
\emph{Rep}: repeat sessions; \emph{Non-Rep}: non-repeat sessions.
}
\end{minipage}
\end{table}

Generally, RepeatNet with repeat outperforms RepeatNet without repeat on all datasets, as shown in Table~\ref{results}.
The results of RepeatNet (with and without repeat) on repeated and non-repeated sessions are shown in Table~\ref{repeat_analysis1} and \ref{repeat_analysis2}.
We can see that the improvements of RepeatNet mainly come from repeated sessions.
Especially on DIGINTICA and LASTFM, RepeatNet improves by 33.91\% and 24.16\% respectively in terms of Recall@20 on repeated sessions.
However, RepeatNet drops a little on non-repeated sessions.
The results indicate that RepeatNet has more potential by explicitly modeling \textit{repeat mode} and \textit{explore mode}. 
But it also shows the limitation of RepeatNet that it seems inclined to repeat recommendations too much if we let it learn the mode probabilities totally from data.
A mechanism should be added to incorporate prior knowledge.

\begin{table}[h]
\centering
\caption{Recall@20 (\%) of RepeatNet (with and without repeat mechanism) on repeat and non-repeat sessions.}
\label{repeat_analysis2}
\begin{tabular}{llcc}
\toprule
\multicolumn{2}{c}{\bf RepeatNet}     & \bf With repeat & \bf No repeat \\ \midrule
\multirow{2}{*}{YOOCHOOSE} & Rep     &      \textbf{97.41}     &      93.70        \\  
                           & Non-Rep &      61.32     &       \textbf{61.95}       \\ \midrule
\multirow{2}{*}{DIGINTICA} & Rep     &    \textbf{99.09}     &      65.18       \\  
                           & Non-Rep &    34.58    &       \textbf{36.73}       \\ \midrule
\multirow{2}{*}{LASTFM}    & Rep     &     \textbf{91.22}      &       67.06       \\  
                           & Non-Rep &     16.79      &     \textbf{20.10}        \\ \bottomrule
\end{tabular}
\end{table}

\subsection{Analysis of the attention vs repeat mechanism}

Neural attention has shown its potential on many tasks \citep{Bahdanau2015,Ren:2017:LCS:3077136.3080792,Li:2017:CAN:3097983.3098115} and also on recommender systems recently \citep{Li:2017:NAS:3132847.3132926,DBLP:conf/ijcai/XiaoY0ZWC17,Chen:2017:ACF:3077136.3080797}.
We compare the results of RepeatNet with and without attention, with and without repeat in Table~\ref{att_analysis1} and \ref{att_analysis2}.
The results show that both repeat and attention mechanisms can improve the results over Improved-GRU4REC.
Importantly, the contributions of attention and repeat mechanisms are complementary as the combination brings further improvements, on all metrics and datasets, demonstrating the need for both.
Besides, we can see that the attention mechanism helps to improve Recall while the repeat mechanism helps to improve MRR.

\begin{table}[h]
\centering
\caption{MRR@20 (\%) of RepeatNet with attention vs with repeat.}
\label{att_analysis1}
\begin{tabular}{@{~}l@{~}c@{~~}c@{~~}c@{~}}
\toprule
\bf RepeatNet         & \bf YOOCHOOSE & \bf DIGINTICA & \bf LASTFM \\ \midrule
No attention &     28.65    &     16.03   &    11.10    \\ 
No repeat    &     30.76    &    13.52     &   10.47   \\
With both         &    \textbf{31.03}     &     \textbf{17.66}      &     \textbf{12.03}   \\ \bottomrule
\end{tabular}
\end{table}

\begin{table}[h]
\centering
\caption{Recall@20 (\%) of RepeatNet with attention vs with repeat.}
\label{att_analysis2}
\begin{tabular}{@{~}l@{~}c@{~~}c@{~~}c@{~}}
\toprule
\bf RepeatNet         & \bf YOOCHOOSE & \bf DIGINTICA & \bf LASTFM \\ \midrule
No attention &    67.74    &     36.50    &   29.47   \\ 
No repeat    &     70.21    &     42.56    &   29.96   \\
With both         &    \textbf{70.71}     &     \textbf{47.79}    &    \textbf{32.38}   \\ \bottomrule
\end{tabular}
\end{table}

\subsection{Analysis of joint learning}

Interestingly, if we jointly train the recommendation loss $L_{rec}$ and the mode prediction probability $L_{mode}$, the overall performance drops a little, as shown in Table~\ref{joint_learning}.
We believe that this is due to the following.
First, $L_{rec}$ is already a good supervisor for learning the mode prediction. 
This conclusion can be drawn from Table~\ref{repeat_analysis1} and \ref{repeat_analysis2} where it shows that RepeatNet (with $L_{rec}$ only) achieves large improvements on repeated sessions.
And the room left for improvement on repeated sessions is relatively small.
Second, RepeatNet (with $L_{rec}$ only) is inclined to repeat recommendation.
Adding $L_{mode}$ further exacerbates the situation.
Besides, $L_{mode}$ assumes that if the next item exists in $I_S$, then it is generated in \textit{repeat mode}, which is not always reasonable.

\begin{table}[h]
\centering
\caption{MRR@20 and Recall@20 (\%) of RepeatNet with and without joint learning.}
\label{joint_learning}
\begin{tabular}{p{2.5cm}cccc}
\toprule
\multirow{2}{*}{\bf Loss} & \multicolumn{2}{c}{\bf YOOCHOOSE} & \multicolumn{2}{c}{\bf LASTFM} \\
\cmidrule(r){2-3}
\cmidrule(r){4-5}
                         & MRR       & Recall      & MRR     & Recall     \\
\midrule                         
$L_{rec}$        &       \textbf{31.03}       &        \textbf{70.71}        &      \textbf{12.03}      &        \textbf{32.38}       \\
$L_{rec}+L_{mode}$           &       28.99      &         69.64        &      11.58      &        31.94     \\
\bottomrule
\end{tabular}
\end{table}


\section{Conclusion and Future Work}
We propose RepeatNet with an encoder-decoder architecture to address \textit{repeat consumption} in the session-based recommendation task.
By incorporating a \textit{repeat-explore mechanism} into \acp{RNN}, RepeatNet can better capture the repeat-or-explore recommendation intent in a session. 
We conduct extensive experiments and analyses on three datasets and demonstrate that RepeatNet outperforms state-of-the-art methods in terms of MRR and Recall. 

RepeatNet can be advanced and extended in several directions.
First, prior knowledge of people can be incorporated to influence \textit{repeat-explore mechanism}. 
Second, more information (e.g., metadata, text) and more factors (e.g., collaborative filtering) can be considered to further improve the performance.
Besides, variants of RepeatNet can be applied to other recommendation tasks, such as content based recommendations.


\section*{Acknowledgments}
This work is supported by the Natural Science Foundation of China (61672324, 61672322), the Natural Science Foundation of Shandong province (2016ZRE27468), the Fundamental Research Funds of Shandong University,
Ahold Delhaize,
Amsterdam Data Science,
the Bloomberg Research Grant program,
Elsevier,
the European Community's Seventh Framework Programme (FP7/2007-2013) under
grant agreement nr 312827 (VOX-Pol),
the Google Faculty Research Awards program,
the Microsoft Research Ph.D. program,
the Netherlands Institute for Sound and Vision,
the Netherlands Organisation for Scientific Research (NWO)
under project nrs
CI-14-25,
652.\-002.\-001,
612.\-001.\-551,
652.\-001.\-003,
and
Yandex.
All content represents the opinion of the authors, which is not necessarily shared or endorsed by their respective employers and/or sponsors.

\subsection*{Code}
To facilitate reproducibility of the results in this paper, we are sharing the code used to run the experiments in this paper at \url{https://github.com/PengjieRen/RepeatNet}.

\bibliography{aaai}

\begin{thebibliography}{}

\bibitem[\protect\citeauthoryear{Anderson \bgroup et al\mbox.\egroup
  }{2014}]{Anderson:2014:DRC:2566486.2568018}
Anderson, A.; Kumar, R.; Tomkins, A.; and Vassilvitskii, S.
\newblock 2014.
\newblock The dynamics of repeat consumption.
\newblock In {\em WWW '14}.

\bibitem[\protect\citeauthoryear{Bahdanau, Cho, and
  Bengio}{2015}]{Bahdanau2015}
Bahdanau, D.; Cho, K.; and Bengio, Y.
\newblock 2015.
\newblock Neural machine translation by jointly learning to align and
  translate.
\newblock In {\em ICLR '15}.

\bibitem[\protect\citeauthoryear{Benson, Kumar, and
  Tomkins}{2016}]{Benson:2016:MUC:2872427.2883024}
Benson, A.~R.; Kumar, R.; and Tomkins, A.
\newblock 2016.
\newblock Modeling user consumption sequences.
\newblock In {\em WWW '16}.

\bibitem[\protect\citeauthoryear{Celma}{2010}]{Celma:Springer2010}
Celma, {\`{O}}.
\newblock 2010.
\newblock {\em Music Recommendation and Discovery in the Long Tail}.
\newblock Springer.

\bibitem[\protect\citeauthoryear{Chen \bgroup et al\mbox.\egroup
  }{2012}]{Chen:2012:PPV:2339530.2339643}
Chen, S.; Moore, J.~L.; Turnbull, D.; and Joachims, T.
\newblock 2012.
\newblock Playlist prediction via metric embedding.
\newblock In {\em KDD '12}.

\bibitem[\protect\citeauthoryear{Chen \bgroup et al\mbox.\egroup
  }{2017}]{Chen:2017:ACF:3077136.3080797}
Chen, J.; Zhang, H.; He, X.; Nie, L.; Liu, W.; and Chua, T.-S.
\newblock 2017.
\newblock Attentive collaborative filtering: Multimedia recommendation with
  item- and component-level attention.
\newblock In {\em SIGIR '17}.

\bibitem[\protect\citeauthoryear{Chen, Wang, and
  Wang}{2015}]{Chen:2015:YRN:2887007.2887011}
Chen, J.; Wang, C.; and Wang, J.
\newblock 2015.
\newblock Will you ``reconsume'' the near past? {Fast} prediction on short-term
  reconsumption behaviors.
\newblock In {\em AAAI '15}.

\bibitem[\protect\citeauthoryear{Cheng \bgroup et al\mbox.\egroup
  }{2017}]{ijcai2017-511}
Cheng, Z.; Shen, J.; Zhu, L.; Kankanhalli, M.; and Nie, L.
\newblock 2017.
\newblock Exploiting music play sequence for music recommendation.
\newblock In {\em IJCAI '17},  3654--3660.

\bibitem[\protect\citeauthoryear{Cheng \bgroup et al\mbox.\egroup
  }{2018}]{Cheng:2018:ALF:3178876.3186145}
Cheng, Z.; Ding, Y.; Zhu, L.; and Kankanhalli, M.
\newblock 2018.
\newblock Aspect-aware latent factor model: Rating prediction with ratings and
  reviews.
\newblock In {\em WWW '18},  639--648.

\bibitem[\protect\citeauthoryear{Davidson \bgroup et al\mbox.\egroup
  }{2010}]{Davidson:2010:YVR:1864708.1864770}
Davidson, J.; Liebald, B.; Liu, J.; Nandy, P.; Van~Vleet, T.; Gargi, U.; Gupta,
  S.; He, Y.; Lambert, M.; Livingston, B.; and Sampath, D.
\newblock 2010.
\newblock The {YouTube} video recommendation system.
\newblock In {\em RecSys '10}.

\bibitem[\protect\citeauthoryear{Glorot and
  Bengio}{2010}]{Glorot2010Understanding}
Glorot, X., and Bengio, Y.
\newblock 2010.
\newblock Understanding the difficulty of training deep feedforward neural
  networks.
\newblock {\em JMLR} 9:249--256.

\bibitem[\protect\citeauthoryear{Gu \bgroup et al\mbox.\egroup
  }{2016}]{gu-EtAl:2016:P16-1}
Gu, J.; Lu, Z.; Li, H.; and Li, V.~O.
\newblock 2016.
\newblock Incorporating copying mechanism in sequence-to-sequence learning.
\newblock In {\em ACL '16}.

\bibitem[\protect\citeauthoryear{He \bgroup et al\mbox.\egroup
  }{2017}]{He:2017:NCF:3038912.3052569}
He, X.; Liao, L.; Zhang, H.; Nie, L.; Hu, X.; and Chua, T.-S.
\newblock 2017.
\newblock Neural collaborative filtering.
\newblock In {\em WWW '17}.

\bibitem[\protect\citeauthoryear{He \bgroup et al\mbox.\egroup
  }{2018a}]{He:2018:APR:3209978.3209981}
He, X.; He, Z.; Du, X.; and Chua, T.-S.
\newblock 2018a.
\newblock Adversarial personalized ranking for recommendation.
\newblock In {\em SIGIR '18},  355--364.

\bibitem[\protect\citeauthoryear{He \bgroup et al\mbox.\egroup
  }{2018b}]{8352808}
He, X.; He, Z.; Song, J.; Liu, Z.; Jiang, Y.-G.; and Chua, T.-S.
\newblock 2018b.
\newblock Nais: Neural attentive item similarity model for recommendation.
\newblock {\em TKDE} 30(12):2354--2366.

\bibitem[\protect\citeauthoryear{Hidasi and
  Karatzoglou}{2017}]{DBLP:journals/corr/HidasiK17}
Hidasi, B., and Karatzoglou, A.
\newblock 2017.
\newblock Recurrent neural networks with top-k gains for session-based
  recommendations.
\newblock {\em CoRR} abs/1706.03847.

\bibitem[\protect\citeauthoryear{Hidasi \bgroup et al\mbox.\egroup
  }{2016a}]{Hidasi2016}
Hidasi, B.; Karatzoglou, A.; Baltrunas, L.; and Tikk, D.
\newblock 2016a.
\newblock Session-based recommendations with recurrent neural networks.
\newblock In {\em ICLR '16}.

\bibitem[\protect\citeauthoryear{Hidasi \bgroup et al\mbox.\egroup
  }{2016b}]{Hidasi:2016:PRN:2959100.2959167}
Hidasi, B.; Quadrana, M.; Karatzoglou, A.; and Tikk, D.
\newblock 2016b.
\newblock Parallel recurrent neural network architectures for feature-rich
  session-based recommendations.
\newblock In {\em RecSys '16}.

\bibitem[\protect\citeauthoryear{Jannach and
  Ludewig}{2017}]{Jannach:2017:RNN:3109859.3109872}
Jannach, D., and Ludewig, M.
\newblock 2017.
\newblock When recurrent neural networks meet the neighborhood for
  session-based recommendation.
\newblock In {\em RecSys '17}.

\bibitem[\protect\citeauthoryear{Lerche, Jannach, and
  Ludewig}{2016}]{Lerche:2016:VRW:2930238.2930244}
Lerche, L.; Jannach, D.; and Ludewig, M.
\newblock 2016.
\newblock On the value of reminders within e-commerce recommendations.
\newblock In {\em UMAP '16}.

\bibitem[\protect\citeauthoryear{Li \bgroup et al\mbox.\egroup
  }{2017a}]{Li:2017:CAN:3097983.3098115}
Li, H.; Min, M.~R.; Ge, Y.; and Kadav, A.
\newblock 2017a.
\newblock A context-aware attention network for interactive question answering.
\newblock In {\em KDD '17}.

\bibitem[\protect\citeauthoryear{Li \bgroup et al\mbox.\egroup
  }{2017b}]{Li:2017:NAS:3132847.3132926}
Li, J.; Ren, P.; Chen, Z.; Ren, Z.; Lian, T.; and Ma, J.
\newblock 2017b.
\newblock Neural attentive session-based recommendation.
\newblock In {\em CIKM '17}.

\bibitem[\protect\citeauthoryear{Mei \bgroup et al\mbox.\egroup
  }{2018}]{Mei:2018:AIN:3269206.3271813}
Mei, L.; Ren, P.; Chen, Z.; Nie, L.; Ma, J.; and Nie, J.-Y.
\newblock 2018.
\newblock An attentive interaction network for context-aware recommendations.
\newblock In {\em CIKM '18},  157--166.

\bibitem[\protect\citeauthoryear{Mobasher \bgroup et al\mbox.\egroup
  }{2002}]{Mobasher:2002:USN:844380.844798}
Mobasher, B.; Dai, H.; Luo, T.; and Nakagawa, M.
\newblock 2002.
\newblock Using sequential and non-sequential patterns in predictive web usage
  mining tasks.
\newblock In {\em ICDM '02}.

\bibitem[\protect\citeauthoryear{Pascanu, Mikolov, and
  Bengio}{2013}]{Pascanu2013}
Pascanu, R.; Mikolov, T.; and Bengio, Y.
\newblock 2013.
\newblock On the difficulty of training recurrent neural networks.
\newblock In {\em ICML '13}.

\bibitem[\protect\citeauthoryear{Quadrana \bgroup et al\mbox.\egroup
  }{2017}]{Quadrana:2017:PSR:3109859.3109896}
Quadrana, M.; Karatzoglou, A.; Hidasi, B.; and Cremonesi, P.
\newblock 2017.
\newblock Personalizing session-based recommendations with hierarchical
  recurrent neural networks.
\newblock In {\em RecSys '17}.

\bibitem[\protect\citeauthoryear{Ren \bgroup et al\mbox.\egroup
  }{2017}]{Ren:2017:LCS:3077136.3080792}
Ren, P.; Chen, Z.; Ren, Z.; Wei, F.; Ma, J.; and de~Rijke, M.
\newblock 2017.
\newblock Leveraging contextual sentence relations for extractive summarization
  using a neural attention model.
\newblock In {\em SIGIR '17}.

\bibitem[\protect\citeauthoryear{Rendle \bgroup et al\mbox.\egroup
  }{2009}]{Rendle:2009:BBP:1795114.1795167}
Rendle, S.; Freudenthaler, C.; Gantner, Z.; and Schmidt-Thieme, L.
\newblock 2009.
\newblock {BPR}: {Bayesian} personalized ranking from implicit feedback.
\newblock In {\em UAI '09}.

\bibitem[\protect\citeauthoryear{Rendle, Freudenthaler, and
  Schmidt-Thieme}{2010}]{Rendle:2010:FPM:1772690.1772773}
Rendle, S.; Freudenthaler, C.; and Schmidt-Thieme, L.
\newblock 2010.
\newblock Factorizing personalized markov chains for next-basket
  recommendation.
\newblock In {\em WWW '10}.

\bibitem[\protect\citeauthoryear{Shani, Heckerman, and
  Brafman}{2005}]{Shani:2005:MRS:1046920.1088715}
Shani, G.; Heckerman, D.; and Brafman, R.~I.
\newblock 2005.
\newblock An {MDP}-based recommender system.
\newblock {\em JMLR} 6:1265--1295.

\bibitem[\protect\citeauthoryear{Srivastava \bgroup et al\mbox.\egroup
  }{2014}]{Srivastava2014}
Srivastava, N.; Hinton, G.; Krizhevsky, A.; Sutskever, I.; and Salakhutdinov,
  R.
\newblock 2014.
\newblock Dropout: A simple way to prevent neural networks from overfitting.
\newblock {\em JMLR} 15(1):1929--1958.

\bibitem[\protect\citeauthoryear{Tan, Xu, and
  Liu}{2016}]{Tan:2016:IRN:2988450.2988452}
Tan, Y.~K.; Xu, X.; and Liu, Y.
\newblock 2016.
\newblock Improved recurrent neural networks for session-based recommendations.
\newblock In {\em DLRS '16}.

\bibitem[\protect\citeauthoryear{Tokui \bgroup et al\mbox.\egroup
  }{2015}]{chainer_learningsys2015}
Tokui, S.; Oono, K.; Hido, S.; and Clayton, J.
\newblock 2015.
\newblock Chainer: a next-generation open source framework for deep learning.
\newblock In {\em NIPS '15}.

\bibitem[\protect\citeauthoryear{Xiao \bgroup et al\mbox.\egroup
  }{2017}]{DBLP:conf/ijcai/XiaoY0ZWC17}
Xiao, J.; Ye, H.; He, X.; Zhang, H.; Wu, F.; and Chua, T.
\newblock 2017.
\newblock Attentional factorization machines: Learning the weight of feature
  interactions via attention networks.
\newblock In {\em IJCAI '17}.

\bibitem[\protect\citeauthoryear{Yap, Li, and
  Yu}{2012}]{Yap:2012:ENR:2260963.2260969}
Yap, G.-E.; Li, X.-L.; and Yu, P.~S.
\newblock 2012.
\newblock Effective next-items recommendation via personalized sequential
  pattern mining.
\newblock In {\em DASFAA '12}.

\bibitem[\protect\citeauthoryear{Zhang \bgroup et al\mbox.\egroup
  }{2014}]{Zhang:2014:SCP:2893873.2894086}
Zhang, Y.; Dai, H.; Xu, C.; Feng, J.; Wang, T.; Bian, J.; Wang, B.; and Liu,
  T.-Y.
\newblock 2014.
\newblock Sequential click prediction for sponsored search with recurrent
  neural networks.
\newblock In {\em AAAI '14}.

\bibitem[\protect\citeauthoryear{Zimdars, Chickering, and
  Meek}{2001}]{Zimdars:2001:UTD:647235.720264}
Zimdars, A.; Chickering, D.~M.; and Meek, C.
\newblock 2001.
\newblock Using temporal data for making recommendations.
\newblock In {\em UAI '01}.

\end{thebibliography}
\bibliographystyle{aaai}

\end{document}